\documentclass[aps, prl, 10pt, superscriptaddress, twocolumn]{revtex4-1}

\usepackage{amsmath,amssymb,mathrsfs}
\usepackage{graphicx}

\newcommand{\la}{\label}
\newcommand{\bbm}{\begin{multline}}
\newcommand{\eem}{\end{multline}}
\newcommand{\be}{\begin{equation}}
\newcommand{\ee}{\end{equation}}
\newcommand{\bea}{\begin{eqnarray}}
\newcommand{\eea}{\end{eqnarray}}
\newcommand{\p}{\partial}

\newcommand{\cE} {\mathcal{E}}

\newcommand{\gzb} {g_{z\bar{z}}}

\newcommand{\comment}[1]{}
\usepackage{color}
\def\red{\color{red}}

\allowdisplaybreaks[1]

\begin{document}

\title{Density-curvature response and gravitational anomaly}

\author{Andrey~Gromov}
\affiliation{Department of Physics and Astronomy, Stony Brook University,  Stony Brook, NY 11794, USA}

\author{Alexander G.~Abanov}
\affiliation{Department of Physics and Astronomy and Simons Center for Geometry and Physics,
Stony Brook University,  Stony Brook, NY 11794, USA}

\date{\today}

\begin{abstract}
We study constraints imposed by the Galilean invariance on linear electromagnetic and elastic responses of two-dimensional gapped systems in background magnetic field.
Exact relations between response functions following from the Ward identities are derived. In addition to viscosity-conductivity relations known in literature we find new relations between the density-curvature response and the chiral central charge.
\end{abstract}



\maketitle


\paragraph{1.Introduction.}Strongly interacting two-dimensional electron gas in magnetic field is notorious for defying analytical approaches. Recently some progress was achieved in understanding these systems with the use of local Galilean symmetry\cite{2012-HoyosSon,son2013newton,Haackdiffeo}. This symmetry is present in simplest models of non-interacting electrons. It is also possible to add non-trivial interactions to this model preserving the local Galilean invariance (LGI) \cite{2012-HoyosSon}. Thus, locally Galilean invariant systems may serve, at the very least, as toy models for FQHE states.

\par In this paper we find the constraints on linear response functions imposed by LGI. In addition to the well known electromagnetic responses we include responses to an external gravitational field. The latter can be used to compute various visco-elastic responses. For systems in background magnetic field some of these constraints were found in Ref.~\onlinecite{2012-HoyosSon}. Those constraints relate wavevector dependent Hall conductivity with Hall viscosity. Later more general relations of this type were obtained \cite{bradlyn-read-2012kubo}.
The relations between various linear response functions derived in this work include the generalization of viscosity-conductivity relation to arbitrary gyromagnetic ratio $g_s$, Kohn's theorem for electric susceptibility and its gravitational analogues. In general, Ward identities impose an infinite number of constraints on the coefficients in wavevector and frequency expansions of response functions.

\par It is well known that the Hamiltonian of a charged particle with gyromagnetic ratio $g_s=2$ is factorizable and has a macroscopic degeneracy of the first Landau level even in the presence of inhomogeneous magnetic field and spatial curvature \cite{aharonov-casher,Alicki-regularity}. In Refs.~\onlinecite{Alicki-regularity,son2013newton} it was argued that at this special value of $g_{s}$ the response functions are regular in the limit of the cyclotron frequency going to infinity. We use this argument together with LGI  to relate the chiral central charge to a bulk density-curvature response. This relation allows us to predict the value of this bulk response for states described by a $K$-matrix  with $\nu\leq 1$. A similar relation for the Laughlin's functions was  recently found in \cite{wiegmann2014}.

\paragraph{2. Galilean invariance.} 
In Galilean invariant systems with one species of particles or with multiple species with equal $\frac{e}{m}$ for each particle there is a relation between the mass current $T^{0i}$ and electric current $J^i$ (see e.g., \cite{1989-GreiterWilczekWitten})
\be \la{current}
	J^i = \frac{e}{m} T^{0i} \,.
\ee
Here we consider the local version of the Galilean symmetry following Ref.~\onlinecite{2006-SonWingate}. The expectation values of the electric current $J^i$ and the stress tensor $T^{ij}$ in the background e/m and gravitational fields can be computed as
\bea
	J^i &=&\frac{1}{\sqrt{g}} \frac{\delta S_{eff}}{\delta A_i}\,,
 	\qquad
	T^{ij} =-\frac{2}{\sqrt{g}} \frac{\delta S_{eff}}{\delta g_{ij}}\,.
 \la{Teff}
\eea
Here $S_{eff}$ is the effective action that captures low energy dynamics of the underlying microscopic theory. The local symmetry of the action that insures (\ref{current}) is \cite{2006-SonWingate}
\bea 
	\delta A_i &=& -\xi^kF_{ki} -mg_{ik}\dot{\xi}^k - \partial_i (\alpha + A^k\xi_k) \,,
 \nonumber \\
	\delta A_0 &=& -\xi^kF_{k0} 
	- \partial_0 (\alpha + A^k\xi_k) + \frac{g_s}{4}\frac{\epsilon^{ij}}{\sqrt{g}}\p_i(g_{jk}\dot{\xi}^k) \,, 
  \la{Gal} \\
	\delta g_{mn} &=& -\xi^k\partial_k g_{mn} - g_{mk}\partial_n\xi^k- g_{nk}\partial_m\xi^k \,,
 \nonumber
\eea
where $F_{ik} = \p_i A_k - \p_k A_i$ is the field strength tensor. The last term in (\ref{Gal}) accounts for the effective magnetic moment of electrons equal $g_{s}/4$. Here and in the following we assume $e=\hbar=c=1$.

These transformations combine a local version of Galilean transformations parameterized by $\xi^{k}(x,t)$ and gauge transformations $\alpha(x,t)$. In the following we use Galilean transformations accompanied by a particular gauge transformation $\alpha = - A_k\xi^k$, so that (\ref{Gal}) have an explicitly gauge invariant form. Conventional (global) Galilean transformations corresponding to a constant velocity $v^{k}$ are given by $\xi^k(x,t) = v^k t$. 

\paragraph{3. Background field separation.}
In the following we assume that the background e/m and gravitational fields are small and smooth deviations from the constant background magnetic field $\overline{F}_{12}=B_{0}$ and flat metric $\overline{g}_{ik}=\delta_{ik}$.  The vector potential can be written as $\overline{A}_i + A_i$, where the first term corresponds to the constant magnetic field. The second term generates time-dependent, inhomogeneous electro-magnetic fields. 

The constant part of the external magnetic field $B_0$ is a parameter of the macroscopic theory and will enter the coefficients in the gradient expansion of the effective action. We do not transform it under Galilean transformations but instead absorb the corresponding part into the transformation laws of the vector potential $A_i$ (compare to eq.~\ref{Gal}) as 
\bea 
 \la{BGtransform}
	\delta A_i &=& -\xi^k\overline{F}_{ki}-\xi^kF_{ki} -mg_{ik}\dot{\xi}^k  \,.
\eea
The external metric is a small perturbation over flat background $g_{ik} = \delta_{ik} + \delta g_{ik}$. 

As we are interested only in the linear response we expand the transformations (\ref{Gal},\ref{BGtransform}) in $A_{\mu}$ and $\delta g_{ik}$ and keep only the terms of zeroth order in fields
\footnote{We have to keep the terms of the first order in fields when analyzing the transformations of the terms of the effective action linear in fields.} 
\bea
	\delta^{(0)} A_i &=& -\epsilon_{ki}\xi^kB_0 -m\dot{\xi}_i \,,
 \nonumber \\
	\delta^{(0)} A_0 &=&  \frac{g_s}{4}\epsilon^{ij}\p_i\dot{\xi}_j \,, 
 \la{gal}\\
	\delta^{(0)} g_{ik} &=&  -\partial_i\xi_k - \partial_k\xi_i \,.
 \nonumber
\eea



\paragraph{4. Building blocks for quadratic effective action.} In the following we plan to use the rotational invariance, locality, gauge invariance and similarities between electro-magnetism and Newton-Cartan gravity to restrict the form of the effective action.

The gauge invariance requires that the effective action depends on the vector potential $A_\mu$ only through electric field $E_i$ and magnetic field $B$. The only exception is the Chern-Simons term which is gauge invariant only up to boundary terms. We also assume that the system under consideration is gapped. Therefore, linear response functions are local, i.e., can be written as Taylor series in frequency and momentum.

We can analyze the gravitational terms in a similar way by introducing an Abelian gauge field that encodes coupling to the background curvature. This field is a non-relativistic (Newton-Cartan) spin-connection \cite{2012-HoyosSon}
\bea
	  \omega_0 &=& -\frac{1}{2} \epsilon^{ab} e^{aj}\p_0 e^b_j \,,
 \la{omega0e}\\
	\omega_i &=& -\frac{1}{2} \epsilon^{ab} e^{aj}\p_i e^b_j 
	- \frac{1}{2\sqrt{g}} \epsilon^{jk} \p_j g_{ik} \,,
\eea
where $e^a_j$ are the time-dependent zwiebeins. This spin connection transforms as an Abelian gauge field under local $SO(2)$ spatial rotations $\omega_{\mu}\to \omega_{\mu}+\p_{\mu}\alpha$ and, therefore, depend only on the metric. 



With the spin connection at hand we construct  the gravi-electric $\cE_i =\p_0 \omega_i - \p_i \omega_0$ and gravi-magnetic  $\frac{1}{2}\sqrt{g}R = \p_1\omega_2 - \p_2\omega_1$ which are explicitly invariant under local $SO(2)$ rotations. Despite the similarities there are some differences between these elastic fields and electro-magnetic ones. First, we consider the metric, not spin connection, as a fundamental field and, therefore, $\cE_{i}$ and $R$ have one more derivative in their definition compared to their electro-magnetic cousins. Second, the parity properties are different: $R$ is a scalar while $B$ is a pseudo-scalar and $\cE_i$ is an axial vector.

In the linear order in deviations from the flat background we have explicitly
\bea
	R &\approx& \p_i \p_j  g_{ij} - \Delta  g_{i}^i \,,
 \\
	\cE_i &\approx& - \frac{1}{2} \epsilon^{jk} \p_j \dot{g}_{ik} \,.
\eea
The spin connection $\omega$ can be expressed in terms of perturbations of the metric as follows.
\bea
	 \omega_0 &=& \frac{1}{2} \epsilon^{jk} \delta g_{ij}\dot{g}_{ik}\, \qquad \omega_i=-\frac{1}{2}\epsilon^{jk}\partial_j \delta g_{ik}\, .
 \la{omegag}
\eea
Notice that the expansion of the zeroth component of the spin connection (\ref{omega0e}) starts from the second order in metric $\delta g$.
There is an additional building block describing dilatations - the trace of the metric which we denote as
\bea
	G \equiv \delta g_{i}^i\,.
\eea

\paragraph{5. Effective action.}
In the following we present the quadratic effective action as a sum
\bea
	S_{eff} = S_{eff}^{(1)}+S_{eff}^{(\eta)}+S_{eff}^{(geom)}+S^{(em)}_{eff}+S^{(g)}_{eff}+S^{(mix)}_{eff}\,.
 \nonumber
\eea
The first contribution collects all ``linear'' terms
\bea
	S_{eff}^{(1)} &=& \int d^2x dt \sqrt{g} \left( -\epsilon_0 + \rho_0 A_0 \right) \,.
 \la{Seff1}
\eea
Notice, that although (\ref{Seff1}) is linear in $A_0$ it also contains (through $\sqrt{g}$) terms quadratic in deviations from the constant background. This term encodes the properties of the unperturbed ground state: energy density $\epsilon_0$ and density $\rho_0 = \frac{\nu}{2\pi l^2}$, where $l^{2}=1/B_{0}$ is the magnetic length and $\nu$ is the filling fraction. 
 
The coefficients in (\ref{Seff1}) and below generally depend on the external magnetic field $B_0$, filling fraction $\nu$ and other microscopic parameters of the system such as the Coulomb gap, cyclotron mass, etc.

The next term has a form
\bea
	S_{eff}^{(\eta)} &=& \int d^2x dt\, \eta_H\epsilon^{jk} g_{ij}\dot{g}_{ik} \,,
 \la{Seffeta} 
\eea
where $\eta_H$ is a function of frequency. One can think about $\eta_H(\omega)$ as of frequency dependent Hall viscosity. We notice comparing to (\ref{omegag}) that the term (\ref{Seffeta}) at zero frequency has a form $2\eta_H(0) \omega_0$ which allows to identify $2\eta_H(0)$ as the orbital spin density and $\bar{s}=2\eta_H(0)/\rho_0$ as an average orbital spin per particle. For the conformal block states the latter is given by $2\bar{s} = \mathcal{\nu}^{-1} + 2h_{\psi}$, where $h_{\psi}$ is the conformal weight of the electron operator in the ``neutral'' sector of the CFT   \cite{2009-Read-HallViscosity,2011-ReadRezayi}.

The next contribution contains topological and geometric terms
\bea
	S_{eff}^{(geom)} &=&  \int
	\left(
	\frac{\sigma_H}{2} AdA
	+ S A d \omega
	+C\omega d\omega 
	\right)\,,
 \la{Seffgeom}
\eea
known as the Chern-Simons, Wen-Zee \cite{WenZeeShiftPaper} and the gravitational Chern-Simons terms, respectively. These terms are special as they are invariant with respect to gauge transformations and local rotations only up to full derivatives. Therefore, in the presence of the boundary they are directly related to the boundary theory and are the natural candidates for encoding universal properties. For the coming discussion of the Galilean invariance it is convenient to allow $\sigma_H$, $S$, and $C$ in (\ref{Seffgeom}) to depend on frequency so that they coincide with their conventional values at zero frequency. 

The electro-magnetic response is represented by
\be
	S^{(em)}_{eff} =  \int d^2x dt \left( \epsilon E^2 + \sigma (\p_iE_i) B - \frac{1}{\mu} B^2 \right) \,.
 \la{Semeff}
\ee
Here $\epsilon$ and $\frac{1}{\mu}$ are electric and magnetic susceptibilities and $\sigma$ encodes the gradient correction to the Hall conductivity.

Analogously, we write down gravitational and mixed terms
\bea
	S^{(g)}_{eff} &=&  \int d^2x dt \Big(\epsilon_g \cE^2 + \sigma_g (\p_i\cE_i) R - \frac{1}{\mu_g} R^2 
 \nonumber \\
	&+& \zeta_3 G R + \zeta_4 G (\p_i\cE_i) + \zeta_5 G^2  \Big) \,,
 \la{Sgeff} \\
	S^{(mix)}_{eff} &=& \int d^2x dt \Big(\epsilon_m (E_i\cE_i) +\sigma_{m1} (\p_i E_i) R 
	- \frac{1}{\mu_m} BR 
 \nonumber \\
 	&+& \sigma_{m2} (\p_i\cE_i) B + \zeta_1 G (\p_i E_i) + \zeta_2 G B\Big) \,.
 \la{Smixeff}
\eea
Equations (\ref{Semeff},\ref{Sgeff},\ref{Smixeff}) contain all possible combinations that can enter real, rotationally, gauge and PT invariant quadratic effective action of a gapped system in transverse constant magnetic field. They define $15$ different response coefficients $\epsilon,\sigma,\mu,\ldots$. 
All coefficients in (\ref{Semeff},\ref{Sgeff},\ref{Smixeff}) are understood as functions of $k^{2}$ and even functions of frequency (in the Fourier space).

\paragraph{6. Hall conductivity and orbital spin.} The next step is to derive Ward identities corresponding to LGI. We apply transformations (\ref{gal}) to $S_{eff}^{(\eta)}+S_{eff}^{(geom)}+S^{(em)}_{eff}+S^{(g)}_{eff}+S^{(mix)}_{eff}$ and transformations (\ref{Gal}) expanded to the first order in fields to $S_{eff}^{(1)}$. We demand the invariance of the full effective action under these transformations up to the terms linear in fields.  This requirement imposes constraints on the response functions in all orders of the gradient expansion. In their full generality these relations are not enlightening and will be presented elsewhere. Here we consider only several particular relations.

In addition to the symmetry requirements, we assume that the underlying microscopic system is gapped and demand that the effective action is local, i.e. the response functions are analytic in frequencies and momenta.

We start with the following relations
\bea 
	\sigma_H &=& \frac{\nu}{2\pi} \frac{\omega_c^2}{\omega_c^2-\omega^2} \,,
 \quad
	S = 2\eta_Hl^2 \frac{\omega_c^2}{\omega_c^2-\omega^2}  \, ,
 \la{S}
\eea
where $\omega_{c}=B_{0}/m$ is the cyclotron frequency.
These are the familiar relations for the Hall conductivity and the Wen-Zee shift \cite{bradlyn-read-2012kubo}. Integrating the charge density $J^0$ from (\ref{Teff}) over the curved manifold and using (\ref{S}) we obtain that the shift in the total charge on the curved manifold of the Euler character $\chi$ is given by
\be
Q = \nu N_{\phi} + \nu\bar{s}\chi \,.
\ee


\paragraph{7. Zero momentum.} At zero momentum $k=0$ the Ward identities that are still exact in frequency  read
\bea
	\epsilon(\omega) &=&\frac{\nu}{4 \pi }\frac{  \omega _c}{\omega _c^2-\omega ^2} \,,
 \quad
	\epsilon_m(\omega) 
	= \eta_Hl^2\frac{\omega _c}{\omega _c^2-\omega ^2} \,.
 \la{Kohn2}
\eea
Remarkably, the first relation determines the homogeneous dielectric response function $\epsilon(\omega,k=0)$ completely. The second one relates its ``mixed'' cousin $\epsilon_m(\omega,k=0)$ to the Hall viscosity $\eta_H(\omega)$. The poles at $\omega_c$ reflect the Kohn's theorem. 
The first relation can be found in Ref.~\onlinecite{son2013newton} while the second one is new.

The next relation is the finite frequency version of the Hall viscosity-conductivity relation \cite{2012-HoyosSon}
\bea
	\!\!\!\!\!\!\!\!\!\!\!\!
	\frac{\sigma(\omega)}{l^2} =
	\frac{\omega _c^2(\omega _c^2+\omega ^2)}{\left(\omega_c^2 -\omega^2\right)^2}
	\left(\eta_Hl^2 -\frac{\nu g_s}{16\pi}\right)-\frac{\omega _c^2 }{\omega_c^2 -\omega^2}
	\frac{\mu^{-1}}{\omega_c l^2}. 
 \la{hoyos}
\eea
Here we slightly generalized the relation obtained in \cite{bradlyn-read-2012kubo} by including an arbitrary $g_s$-factor.

We also find two elastic analogues of (\ref{hoyos})  
\bea   
	\!\!\!\!
	\frac{\mu_m^{-1}(\omega)}{\omega_cl^2} 
	&=& \frac{C}{2} - \frac{g_s}{4}\eta_Hl^2
	\frac{\omega_c^2+\omega^2}{\omega_c^2-\omega^2}
	-\frac{\sigma_{m1}}{l^2} +\frac{\epsilon^{(1)}_m\omega^2}{\omega_c} ,
 \la{BR} \\
	\frac{\sigma_{m2}}{l^2} &=& \frac{g_s}{2} \eta_Hl^2 
	\frac{\omega_c^2}{\omega_c^2 - \omega^2} + \left(2\epsilon^{(1)}_m - \epsilon_g\right)\omega_c \,,
  \la{BR1} 
\eea
where we introduced $(kl)^2\epsilon^{(1)}_m  = \epsilon_m(k,\omega) - \epsilon_m(0,\omega)$.

The coefficients $\zeta_1,\ldots,\zeta_5$ are completely fixed by the Galilean invariance in terms of other coefficients. Their expansions start with $\omega^2$ and we do not list them here.

\paragraph{8. Regularity of the limit $\omega_c\rightarrow \infty$.} 
Let us consider the static limit $\omega=0$ of (\ref{BR})
\be 
 \la{Ceta}
 	m\mu^{-1}_m(0) = \frac{C}{2} - \frac{\nu\bar{s}g_s}{16\pi} - \frac{1}{l^2}\sigma_{m1}(0)\,.
\ee
The kinetic coefficient $\sigma_{m1}(0)$ describes the contribution to the expectation value of the density proportional to the Laplacian of curvature $\Delta R$. We introduce $b=-8\pi \sigma_{m1}(0)/l^{2}$ which can be defined as a coefficient in the gradient expansion for the static density-curvature response 
\be
	\delta\rho = \frac{S}{2}R +\frac{b}{8\pi} l^2\Delta R +\ldots\,.
 \la{bdef}
\ee
Here the shift $S=\frac{\nu\bar{s}}{2\pi}$ describes the leading term. The subleading coefficient $b$ was introduced in Ref.~\onlinecite{wiegmann2014} and shown to be related to $(kl)^6$ coefficient in the static structure factor.

For $g_s=2$ the ground state of noninteracting electrons is degenerate even in the presence of inhomogenious background fields and it is expected that the limit $m\to0$ (i.e., $\omega_c\to \infty$) is regular for $\nu\leq1$ \cite{Alicki-regularity,son2013newton}. Therefore, $\mu_{m}^{-1}(0)$ is finite in the limit $m\to 0$ at $g_{s}=2$. 

We take the limit $m\to 0$ of (\ref{Ceta}) at $g_{s}=2$. The left hand site vanishes and we find a relation between the coefficients of the Wen-Zee and gCS terms  (\ref{Seffgeom}) and the coefficient $b$
\be
 \la{gCS}
	C =  \frac{S}{2} - \frac{b}{4\pi} \,.
\ee
This relation is obtained for $g_s=2$. However, $b$ is a response of the density to curvature and cannot depend on $g_s$. Therefore, the relation (\ref{gCS}) is valid for general $g_{s}$.

\paragraph{9. Chiral central charge.} 
Let us split the geometric part of the effective action (\ref{Seffgeom}) as
\be
 \la{cEFF}
	S^{(geom)}_{eff} = \int \frac{\nu}{4\pi} 
	\left(A+\bar{s}\omega\right)d(A+\bar{s}\omega) - \frac{c}{48\pi}\omega d \omega\,.
\ee
Here we used (\ref{S}) at zero frequency.
The first contribution in (\ref{cEFF}) reflects the Wen-Zee argument \cite{WenZeeShiftPaper} (see also \cite{son2013newton}) stating that every electron carries not only charge, but also intrinsic orbital spin $\bar{s}$ that couples to the curvature. Thus, in any transport process the electric current will be accompanied by the ``spin current''. Formally, this amounts to changing the vector potential as $A_i \rightarrow A_i + \bar{s}\omega_i$. We have noted in \cite{abanov2014} however, that even in the noninteracting case with $\nu=1$ there is an additional contribution to gCS term represented by the second term in (\ref{cEFF}). Comparing (\ref{Seffgeom}) with (\ref{cEFF}) we identify $C=\frac{\nu}{4\pi}\bar{s}^{2}-\frac{c}{48\pi}$ and rewrite (\ref{gCS}) as
\be
 \la{b}
	b = \nu\bar{s}(1-\bar{s}) + \frac{c}{12} \,.
\ee
This equation relates the coefficients of geometric terms with the static bulk density-curvature response. A relation of this kind appeared recently in \cite{wiegmann2014}. However, the corresponding relation of \cite{wiegmann2014} needs to be properly generalized in order to be applicable to the states beyond Laughlin's.

We refer to $c$ as to the chiral central charge. In relativistic physics $c$ is related to the gravitational anomaly at the boundary \cite{Stone-Gravitational}.
%

Let us consider the relation (\ref{b}) for few cases where $b$ has been computed independently.
The first such case is non-interacting fermions filling the lowest Landau level $\nu=1$. It was found in \cite{abanov2014} that in this case $\nu =1$, $\bar{s}=\frac{1}{2}$, and $b=8\pi \sigma_{m1}(0)/l^{2}=1/3$.
Then (\ref{b}) gives $c=1$ corresponding to $C=\frac{1}{24\pi}$ and is in agreement with the straightforward calculation of \cite{abanov2014}.

For Laughlin states $\nu \bar{s}=1/2$ and using $b  = \frac{1}{3} + \frac{\nu - 1}{4\nu}$ calculated in \cite{wiegmann2014} we predict  using (\ref{b}) and assuming that the results of \cite{wiegmann2014} are compatible with Galilean invariance
\be
	C =  \frac{1}{8\pi} - \frac{1}{4\pi}b = \frac{1}{24\pi} + \frac{1}{2\pi}\frac{\nu^{-1}-1}{8} 
 \la{CLaugh}
\ee
again corresponding to $c=1$.

In both cases the boundary theory is the chiral boson $c=1$ and the results given by (\ref{b}) are in agreement with our expectations for the (chiral) central charge. Therefore, we conjecture that $c$ in (\ref{cEFF}) coincides with the central charge of boundary theory for all other states of FQHE hierarchy.


Note that the relation (\ref{b}) was derived using regularity conditions at $g_s=2$ specific for $\nu\leq 1$ and is not supposed to hold for $\nu> 1$. 
However, for non-interacting case with $\nu = N$ we found using the results of \cite{abanov2014} that (\ref{b}) can still be written as a sum over filled Landau levels
\be
 \la{beyond LLL}
	b = \sum_{n=1}^N \left(\nu_n\bar{s}_n(1-\bar{s}_n) + \frac{c_n}{12}\right) \,.
\ee
Here $\bar{s}_n =\frac{2n-1}{2}$, $\nu_n=1$ and $c_n=1$ for the $n$-th Landau level.

The significance of the equations (\ref{cEFF},\ref{b}) is that in the non-relativistic case, the averaging over the microscopic degrees of freedom produces two gCS terms. One originates from the coupling of the orbital spin to the curvature and the other one is related to the gravitational anomaly of the boundary.

\paragraph{10. Abelian Quantum Hall states.} For general Abelian states we re-write the geometric action (\ref{cEFF})  as 
\be
 \la{KcEFF}
	S^{(geom)}_{eff} = \frac{1}{4\pi}\int  
	\left(t_iA+\bar{s}_i\omega\right)K^{-1}_{ij} d(t_jA+\bar{s}_j\omega) - \frac{c}{12}\omega d \omega \,,
\ee
where $K$-matrix, charge vector $t_i$ and spin vector $\bar{s}_i$ characterize the state \cite{wen-Kmatrix}.
Then (\ref{gCS}) takes the form (in matrix notations)
\be
	\frac{c}{12} = (\bar{s}-t)^t K^{-1} \bar{s} + b 
 \la{Kb}
\ee
generalizing (\ref{b}) to more general Abelian Quantum Hall states. Here the parameter $c$ counts the number of chiral propagating modes and is equal to $c=n_+-n_-$, where $n_\pm$ is the number of positive/negative eigenvalues of $K$-matrix, respectively.

We conclude this section with few examples of applications of (\ref{Kb}) to some well-known FQHE states.
For the Laughlin's state $\nu=\frac{1}{m}$, $K=(m)$, $t=1$, $\bar{s}=m/2$, $c=1$ and we obtain $b = \frac{1}{3} - \frac{m-1}{4}$.  For the corresponding particle-hole conjugated state $\nu=1-1/m$, $K =\left(\begin{array}{cc}
1 & 1 \\
 1 &1-m\\
\end{array}\right)$,  $t=(1,0)$, $\bar{s}=(\frac{1}{2},\frac{1-m}{2})$,  and $c=0$  \cite{wen-Kmatrix}. The relation (\ref{Kb}) gives $b = \frac{m-1}{4}$. 


As an example of non-Abelian state we consider the fermionic Pfaffian state with 
$\nu=1/2$, $t=(-1,-2)$, $\bar{s}=(-3/2,-3)$, $c=3/2$, and  
$K =\left(\begin{array}{cc}
3 & 4 \\
 4 &8\\
\end{array}\right)$ \cite{wen-Kmatrix,Ardonne}. We obtain $b=-1/4$. 

%

\paragraph{11. Thermal Hall effect.}
It has been demonstrated that the thermal Hall current (the Leduc-Righi effect) is related to the chiral central charge of edge modes via the relation \cite{Kane-Fisher-LR,CappelliLR}
\be
	K_H = \frac{\partial J_H}{\partial T}  = \frac{\pi k_B^2T}{6}c \,.
\ee
We use (\ref{KcEFF}) in order to express the thermal hall conductivity through other response functions.
\be
 	\frac{K_H}{2\pi  k_B^2 T} = (\bar{s}-t)^t K^{-1} \bar{s} + b \,.
 \la{KHb}
\ee
An important remark is in order. Eq.~(\ref{KHb}) allows to obtain the thermal Hall response in terms of the bulk quantities. Of course, ``measuring" $b$ involves gradients of curvature or ``tidal forces'' (c.f., Ref.~\onlinecite{Stone-Gravitational,BradlynRead}).


\paragraph{13. Conclusion.}
In this work we have explored the constraints imposed by the local Galilean invariance on linear electromagnetic and gravitational responses of gapped systems in the background of quantizing magnetic field. Several new relations between linear response functions have been found (see e.g., (\ref{BR},\ref{BR1})). Using the regularity of the limit of large cyclotron frequency $\omega_{c}\to\infty$ in addition to the Galilean invariance, we found a relation (\ref{b},\ref{Kb}) between the bulk density-curvature response coefficient $b$ (\ref{bdef}) and the chiral central charge. The relation has been tested for the cases of non-interacting electrons and for Laughlin's states using the results of \cite{abanov2014,wiegmann2014}. As an application we use the relation to predict the values of the density-curvature response $b$ for several other Quantum Hall states. It would be interesting to understand whether the expression (\ref{Kb}) can be derived without the use of the Galilean invariance.

%

We acknowledge discussions with B. Bradlyn, T. Can, M. Haack, H. Hansson, P. Wiegmann, and M. Zaletel.
 The work of A.G.A. was supported by the NSF under grant no. DMR-1206790.

\bibliographystyle{my-refs}




\bibliography{abanov-bibliography}

\end{document}